\documentclass[12pt,a4paper]{article}
\usepackage{amsmath}
\usepackage[dvips]{graphicx}
\input epsf
\usepackage{times}
\begin{document}
\date{}
\author{S. M. Troshin, N. E. Tyurin\\
\small\it{ Institute for High Energy Physics,}\\
\small\it{ 142280 Protvino, Moscow Region, Russia}}
\title{Angular structure of diffraction dissociation at LHC}
\maketitle

\begin{abstract}
We consider angular dependence of the diffractive dissociation
processes at high energies.  It appeared that angular dependence
of the diffractive dissociation has no dip and  bump structure at
the LHC energies and would show a smooth power-like decrease with
$t$. The normalized differential cross-section has a scaling
behavior and depends  on the ratio $-t/M^2$ only.
These results reflect general trends of unitarity limitations at high
energies and are realized within the chiral quark model with
unitarization  performed through the   generalized reaction matrix.
\end{abstract}

\section*{Introduction}

 During  recent years CERN, DESY and FNAL have been
producing  interesting results on diffractive production in
hadron and deep-inelastic processes \cite{rev}. Discovery of hard
diffraction at CERN S$\bar p p$S and  diffractive events in the
deep-inelastic scattering at HERA were among the most surprising
results obtained recently. Significant fraction of high-$t$
events among the diffractive events  in deep-inelastic scattering and
hadron-hadron interactions were also observed at HERA and Tevatron
respectively. These experimental discoveries
 renewed interest in the experimental and theoretical
  studies of the diffractive
production processes.

 The
understanding of the diffractive interactions plays fundamental
role in the studies of a high-energy limit of the modern strong
interaction theory --- QCD. This field is  most interesting one
since there are no firmly established computational methods within
QCD.
However, the progress   in understanding of high--energy basic
features and the underlying dynamics of diffraction mechanism can
be traced in the recent theoretical papers (cf. review
\cite{hebe}).

It is of  high importance to have the experimental data at
highest possible energy and there is no doubt that soft and
 hard diffractive
processes should be measured at the LHC. Besides elastic scattering
(total and elastic cross-sections) it is important to measure the
inelastic diffractive final states, jet production, hadron
transverse momentum distribution, strange and charmed particle
productions. The single diffraction dissociation is a most simple
inelastic diffractive process and studies of soft and hard final
states in this process could become a next step after elastic
measurements. Such measurements are crucial for  understanding
of the microscopic nature of  driving mechanism called Pomeron,
its possible parton structure, high-energy limit of strong
interactions and  approaching  the asymptotia.

 At the currently available energies there were obtained
several sets of the  experimental data
 for single diffraction production process
\begin{equation}
h_1+h_2\rightarrow h_1+h_2^* \label{bpr}
\end{equation}
when the hadron $h_2$ is excited to the state $h_2^*$ with
invariant mass $M$ and the same quantum numbers. Its subsequent
decay results in the multiparticle final state. The inclusive
differential cross--section shows a simple dependence on the
invariant mass $M$:
\begin{equation}
\frac{d\sigma _{diff}}{dM^2} \propto \frac{1}{M^2}. \label{mdep}
\end{equation}
Meanwhile, energy dependence of the diffractive production
cross--section $\sigma _{diff}(s)$ is not so evident from the
experimental data. The    data obtained at FNAL seem to
demonstrate a growing diffractive cross--section but some data
obtained at CERN may indicate a falling diffractive
cross--section. Resolution of this problem is
important in particular for the study of the role of unitarity in
the inelastic diffraction and its asymptotical properties.

The angular dependence of the diffractive dissociation is even
less clear. Similarity between elastic and inelastic diffraction
in the $t$-channel approach suggests that the latter one would
have similar to elastic scattering behavior of the differential
cross-section. However, it cannot be taken for granted and   e.g.
transverse momentum distribution of diffractive events in the
deep-inelastic scattering at HERA  shows a power-like behavior
with no apparent dips \cite{will}. Similar behavior was observed
also in the hadronic diffraction dissociation process at CERN
\cite{cern} where also no dip and bump structure was observed.
Angular dependence of diffraction dissociation together with the
measurements of the differential cross--section in elastic
scattering would allow to determine the geometrical properties of
elastic and inelastic diffraction, their similar and distinctive
features and  origin.

The  experimental regularities of diffractive production can be
described in the framework of different approaches, e.g. models
for inelastic diffraction based on $s$-channel absorptive
unitarity were considered in  \cite{canes}, optical approach was
used in  \cite{meng}, dipole Pomeron and effective Pomeron with
energy-dependent intercept were applied in Refs. \cite{mart} and
\cite{erh}, respectively. $1/M^2$ dependence is naturally
described by the triple--pomeron diagrams in the framework of
Regge--model \cite{kaid}. The proposed in Ref. \cite{donl}
similarity between the pomeron and photon exchanges allowed to
calculate diffractive dissociation cross--section in terms of
structure function $\nu W_2$  measured in deep inelastic lepton
scattering. The explanation of $M^2$--dependence in the framework
of optical model considering diffractive dissociation as a
bremsstrahlung where virtual quanta are released from a strong
field was made in Ref. \cite{fass}. Explanation of  the smallness
of a large mass diffraction as a result of the strong
non-perturbative interaction of gluons was given in \cite{kopel}.
The list of the above references is certainly far from being
complete and many theoretical papers devoted to the various
aspects of diffractive production were not mentioned.

 In Ref. \cite{ddis} for
the description of single diffractive processes we  used
geometrical notions on quark scattering in approach based on
unitarity for the scattering amplitude and chiral quark model for
hadron structure. Motivated by the importance of this process for
the study of long distance dynamics we have shown how the energy
and $M^2$--dependencies can be obtained in the approach to hadron
interactions with account for the spontaneous breaking of chiral
symmetry and presence of a quark condensate inside a hadron.

In this paper we consider transverse momentum distribution in the
framework of the above approach and show that the diffraction cone
would be suppressed and disappears at LHC energy in the
production process $p+p\to p+X$.
\section{Amplitude of the
diffractive production}

To construct the amplitude of the diffractive production process
let us remind that the unitary equation for the scattering
amplitude
\begin{equation}
\mbox{Im} f(s,b) = |f (s,b)|^2 + \eta (s,b)
\end{equation}
allows one to express the inelastic channel contribution
\begin{equation}
\eta (s,b) = \sum_n \sigma _n(s,b)
\end{equation}
through the $U$-matrix
\begin{equation}
\eta  (s,b) =\mbox{Im} U(s,b) |1-iU(s,b)|^{-2}, \label{xxx}
\end{equation}
and to get respectively the  total inelastic cross-section
\begin{equation}
\sigma _{inel}(s) = 8\pi \int^{\infty}_{0}
bdb\frac{\mbox{Im}U(s,b)}{|1-iU(s,b)|^2}.
\end{equation}
The quantity $\mbox{Im} U(s,b)$ can be represented in the following form
\begin{equation}
\mbox{Im} U(s,b) = \sum_n \bar U_n (s,b),
\end{equation}
where
\begin{equation}
\bar U_n (s,b) = \int d\Gamma_n |U_n(s,b,\{\xi _n\})
\end{equation}
and $d\Gamma_n$ is an element of $n$-particle phase space  volume
and $\{\xi _n\}$ is the set of kinematical variables related to
the $n$--particle final state. Sum in the right hand side of this
equation runs over all  inelastic final  states which include as
well diffractive as non-diffractive ones.  To  obtain the
cross-section  of the diffractive  dissociation  process
(\ref{bpr})   we should single out in this sum the  final  states
corresponding  to  the process (\ref{bpr}) .  Let  for  simplicity
consider  the case  of  pure imaginary $U$-matrix, i.e. $U\to iU$.
Then we can represent $d\sigma _{diff}/dM^2$ in the   form
\begin{equation}
\frac{d\sigma _{diff}}{dM^2} = 8\pi
\int^{\infty}_{0} bdb
\frac{U_{diff}(s,b,M)}{[1+U(s,b)]^2}\label{y}
\end{equation}
where   $U_{diff}(s,b,M)$  includes  contributions
from all the
final states $|n\rangle_{diff}$  which  result  from  the  decay  of  the
excited  hadron   $h^*_2$   of  mass   $M$:
 $h^*_2\rightarrow |n\rangle _{diff}$. The corresponding
  impact parameter amplitude
$F_{diff}(s,b,M)$ can be written in this pure imaginary case as a
square root of the cross-section, i.e.
\begin{equation}\label{ampl}
  F_{diff}(s,b,M)={\sqrt{U_{diff}(s,b,M)}}/{[1+U(s,b)]}
\end{equation}
and the amplitude $F_{diff}(s,t,M)$ is
\begin{equation}\label{amplt}
  F_{diff}(s,t,M)=\frac{is}{\pi^2}\int_0^\infty
 bdb J_0(b\sqrt{-t}){\sqrt{U_{diff}(s,b,M)}}/{[1+U(s,b)]}
  .
\end{equation}

In Refs. \cite{ddis,nuov} we used the notions of effective chiral
quark model
 for the description of elastic
scattering  and diffractive production. Different aspects of
hadron dynamics were accounted in the framework of effective
Lagrangian approach. The picture used for the hadron structure
implies that overlapping  and interaction of peripheral
condensates at hadron collision  occurs  at the first stage. In
the overlapping region the condensates interact and as a result
the massive quarks appear.  Those quarks are transient
ones: they are transformed back into the condensates of the final
hadrons in elastic scattering and the diffraction dissociation processes.
In elastic scattering  each of the constituent valence quarks located in the
central part of the hadron
is supposed in the model to scatter in a quasi-inde\-pen\-dent way by the
produced virtual  quark pairs at given impact parameter and by  the
other valence quarks.

 For consideration of the  diffractive
production  at the quark level we have extended  the  picture  of
hadron  interaction in case of elastic scattering \cite{nuov}.
Since the constituent quark is an extended object there is a
non--zero probability of its excitation at the first stage of
hadron interaction  when peripheral
condensates interact. Therefore it seems rather
natural to assume  that the origin of
diffractive production process is an excitation of one of  the
valence quarks in colliding hadron: $Q\rightarrow Q^*$, its
subsequent scattering and then decay into the final state. The  excited
constituent quark is scattered similar to  other valence quarks in
a quasi-independent way. The function $U_{diff}(s,b,M)$ can be
represented then as a product
\begin{equation}
U_{diff}(s,b,M) = \prod^{N-1}_{Q=1} \langle f_Q(s,b)\rangle\langle
f_{Q^*}(s,b,M_{Q^*})\rangle,
\end{equation}
where $M_{Q^*}$ is the mass of excited constituent quark.
This mass is taken  to be  proportional to the mass $M$ of excited hadron $h_2^*$.
This assumption is based on the
additivity of constituent quark masses in the hadron and the
absence of the diffractive
radiation from the other constituent quarks.

In the model the $b$--dependence of the amplitudes $\langle f_{Q}
\rangle $ and $\langle f_{Q^*} \rangle $
 is related to
formfactor of the constituent quark and excited constituent quark respectively.
 The strong interaction radius of constituent
quark is determined by its mass. We suppose that the same is valid
for the size of excited quark, i.e.
$r_{Q^*} = \xi /M_{Q^*}$. The expression for
$U_{diff}(s,b,M)$ can be rewritten then in the following form:
\begin{equation}
U_{diff}(s,b,M) =g^* U(s,b)\exp[-(M_{Q*}-m_Q)b/\xi ],
\end{equation}
where constant $g^*$ is  proportional to  the
 relative probability
of excitation of the constituent quark.
The value of $g^*$ is a non-zero one, however,
 $g^*<1$ since we expect that the excitation of
 any constituent quark  has
lower probability compared  to the probability for this quark to stay
unexcited.  The excited quark is  unstable and its subsequent
decay is associated with the decay of  hadron  $h_2^*$ into
the multiparticle final state $|n\rangle _{diff}$. The expression
for $U(s,b)$ is the following \cite{ddis}:
\begin{equation}
U(s,b) = g(s)\exp(-\tilde M b/\xi )\equiv \tilde G \left [1+\alpha
\frac{\sqrt{s}}{m_Q}\right]^N \exp(-\tilde M b/\xi ), \label{x}
\end{equation}
where $\tilde M =\sum^N_{Q=1}m_Q$ and $N$ is the total number of
the constituent quarks in  colliding hadrons.

\section{Total and differential cross-sections of the diffractive
production}
 The cross-section of diffractive dissociation process
is given by Eq. (\ref{y}) and has the following $s$ and
$M^2$ dependence
\begin{equation}
\frac{d\sigma_{diff}}{dM^2}\simeq \frac{8\pi  g^*\xi
^2}{(M_{Q^*}-m^2_Q)^2}\eta (s,0)\simeq \frac{8\pi  g^*\xi ^2}{M^2}
\eta(s,0)
\end{equation}
Thus,   we   have a familiar $1/M^2$  dependence  of the
diffraction cross-section   which is related in our model to the
geometrical size  of excited  constituent quark. The energy
dependence of single diffractive cross-section has  the
form
\begin{equation}
\sigma_{diff}(s) = 8\pi  g^*\xi ^2\eta(s,0)
\int^{M^2_1}_{M^2_0}
\frac{dM^2}{M^2} = 8\pi  g^*\xi ^2\eta(s,0)
\ln\frac{s(1-x_1)}{M^2_0},\label{z}
\end{equation}
where $x_1$ is the lower limit of the relative momentum of hadron
$h_1$ and corresponds to the experimental
constraint on diffractive process $x_1\simeq  0.8-0.9$.
 Eq. (\ref{z}) shows   that the   total
cross-section of diffractive dissociation has a non-trivial energy
dependence which is determined by the
 contribution of inelastic  channels  into  unitarity equation at
 zero value of impact parameter. The dependence of $\eta (s,0)$ is
determined  by Eq. (\ref{xxx}), where   $U(s,b)$ is
given by Eq. (\ref{x}).

 At $s\leq s_0$, ($s_0$ is determined by
equation $|U(s_0, 0)|=1$) $\eta(s,0)$ increases with  energy. This
increase as it  follows  from  Eq. (\ref{x})  and from the
experimental   data \cite{mitt} is rather slow one. However at
$s\geq s_0$, $\eta(s,0)$ reaches its maximum value $\eta
(s,0)=1/4$ and at $s > s_0$, the function $\eta(s,0)$ decreases
with energy and at  asymptotical energies the inelastic
diffraction cross section drops to zero. But at the LHC energy
$\sqrt{s}=14$ $GeV$ the single diffractive inelastic
cross-sections  can reach the value of $2.4$ mb  \cite{lhc}
and therefore might be quite significant.

Hence, it worth to consider the structure of the
corresponding angular distribution. The corresponding amplitude
$F_{diff}(s,t,M)$ can be calculated analytically. To do so
 we continue the amplitudes
$F_{diff}(s,\beta, M),\,\beta =b^2$, to the complex
 $\beta $--plane and transform the Fourier--Bessel integral over impact
parameter into the integral in the complex $\beta $ -- plane over
the contour $C$ which goes around the positive semiaxis.
Then  for the  amplitude $F_{diff}(s,t,M)$ the following
representation takes place:
\begin{equation}\label{impl}
F_{diff}(s,t,M)  =  -\frac{is}{2\pi ^3}\int_C d\beta
F_{diff}(s,\beta , M )K_0(\sqrt{t\beta })
\end{equation}
where $K_0(x)$ is the modified Bessel function. The amplitude
 $F_{diff}(s,\beta , M )$ has the poles in the
$\beta $--plane determined by  equation
\begin{equation}\label{poles}
1+U(s,\beta )=0.
\end{equation}
The solutions of this equation can be written as
\begin{equation}
\beta _n(s)=\frac{\xi^2}{\tilde M^2}\,\left\{\,\ln g(s)+\,i \pi
n\,\right\},\, n=\pm 1, \pm 3,\ldots \label{polloc}
\end{equation}
 The amplitude $F_{diff}(s,\beta , M )$ besides the poles has a
  branching point at $\beta =0$.

Therefore the  amplitude $F_{diff}(s,t,M)$  can be represented as
a sum of the pole contribution and the contribution of the cut:
\begin{equation}
F_{diff}(s,t, M)=F_{diff,p}(s,t, M)+F_{diff,c}(s,t, M)\label{sum}
\end{equation}

Up to this point the calculation is similar to the case of elastic
scattering. For elastic scattering amplitude $F(s,t)$ the pole and
cut contributions are decoupled dynamically when $g(s)\rightarrow
\infty  $ at $s\rightarrow \infty  $ \cite{trty}. Contribution of
the poles determines the
 elastic amplitude in the region $|t|/s \ll 1\,(t\neq 0)$.  The
amplitude in this region can be represented in a form of  series
 over the parameter $\tau (\sqrt{-t})$:
\begin{equation}
F(s,t)=s\sum_{k=1}^\infty  \tau ^k(\sqrt{-t})\varphi
_k[R(s),\sqrt{-t}], \label{5.14}
\end{equation}
where $\varphi _k[R(s),\sqrt{-t}]$ are the oscillating functions
of the variable  $\sqrt{-t}$. The parameter $\tau $
 decreases exponentially with $\sqrt{-t}$:
\[
\tau (\sqrt{-t})=\exp (-\frac{2\pi\xi }{\tilde M }\sqrt{-t}).
\]
This series reproduces diffraction peak and familiar dip-bump
structure of the differential cross-section in elastic scattering.
In the region of moderate $t$ it is sufficient to take into
account few or even one of the terms of series Eq. \ref{5.14}. The
differential cross-section in this region has familiar Orear
behavior.

However, the situation is different in the case of diffraction
production. Instead of dynamical separation of the pole and cut
contribution discussed above when
\begin{equation}
 F_p=O(s\ln ^{1/2} g(s)),\quad F_c=O(s[g(s)]^{-1}),
\end{equation}
we have a  suppression of the pole contribution at high
energies since  at fixed $t$
\begin{equation}
F_{diff,p}=O(s[g(s)]^{-\frac {M}{2\tilde {M}}}\ln ^{1/2}
g(s)),\quad F_{diff,c}=O(s[g(s)]^{-\frac {1}{2}}),
\end{equation}
i.e. the pole contribution is suppressed at high energies where
$g(s)>1$ since $M>\tilde M$. Therefore, at all $t$ values   we
will have
\begin{equation}
F_{diff}(s,t, M)\simeq F_{diff,c}(s,t, M),\label{dom}
\end{equation}
where
\begin{equation}
 F_{diff,c}(s,t, M)\simeq ig^*g^{-1/2}(s)(1-\frac{t}{\bar M^2})^{-3/2},
\end{equation}
where $\bar M=(M-\tilde M-1)/2\xi$. This means that the
differential cross-section of the diffraction production will have
smooth dependence on $t$ with no apparent dips and bumps
\begin{equation}\label{dsig}
\frac {d\sigma_{diff}}{dtdM^2}\propto (1-\frac{t}{\bar M^2})^{-3}.
\end{equation}
It is interesting to note that at large values of $M\gg \tilde M$
the normalized differential cross-section
$\frac{1}{\sigma_0}\frac{d\sigma }{ dtdM^2}$ ($\sigma_0$ is the
value of cross-section at $t=0$) will exhibit scaling behavior
\begin{equation}\label{scal}
\frac{1}{\sigma_0}\frac{d\sigma }{dtdM^2}=f(-t/M^2),
\end{equation}
 and explicit form of the function $f(-t/M^2)$ is
the following
\begin{equation}\label{ftau}
 f(-t/M^2)=(1-4\xi ^2t/M^2)^{-3}.
\end{equation}
The numerical parameter $\xi$ in the model is about 2. The
function Eq. (\ref{scal}) against the variable $-t/M^2$ is
depicted in the Fig. 1.
\begin{center}
\begin{figure}[t]
 \begin{center}
\epsfxsize=80 mm  \epsfbox{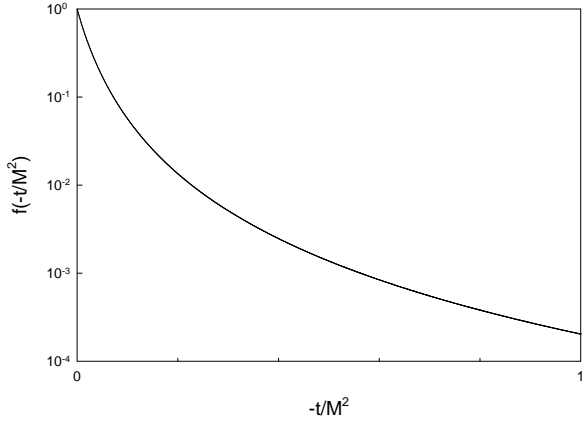}
 \end{center}
\caption{Scaling behavior of the normalized differential
cross-section $\frac{1}{\sigma_0}\frac{d\sigma }{ dtdM^2}$.}
 \end{figure}
\end{center}
 The above scaling has been obtained in the model approach,
however it might have a more general meaning.

 The angular structure of
diffraction dissociation processes given by Eq. (\ref{dsig}) takes
place  at high energies where $g(s)>1$ while at moderate energies
where $g(s)\simeq 1$ the both contributions from poles and cut are
significant. At low energies where $g(s)\ll 1$ the situation is
similar to the elastic scattering, i.e. there  diffraction
cone and possible dip-bump structure should be present
in the region of small values
 of $t$. In this region
\begin{equation}
F_{diff}(s,t, M^2)=s\sum_{k=1}^\infty  \tau ^k(\sqrt{-t})\varphi
_{diff,k}[R(s),\sqrt{-t}, M^2], \label{poldif}
\end{equation}
where the parameter $\tau(\sqrt{-t})$ is the same as  in the
elastic case and
\begin{equation}
\varphi_{diff,k}=O(s[g(s)]^{-\frac {M}{2\tilde {M}}}\ln ^{1/2}
g(s)).
\end{equation}

\section*{Conclusion}
We considered  behavior of the differential cross-section for
single inelastic diffraction. At low and moderate
energies behavior of the differential cross-section
will be rather complicated and incorporate diffraction cone, Orear
type and power-like dependencies.

However, at high energies a
simple power-like dependence on $t$ is predicted. It was shown
that the normalized differential cross-section has a scaling form
and only depends  on the ratio $-t/M^2$ at large values of $M^2$.

In fact, our particular comparative analysis of the poles and
cut contributions has very little with the  model form
of the $U$--matrix.
The approach  is based on the combination of unitarity
with the diffraction production mechanism via  excitation and
subsequent decay of a constituent quark. This is why it may have
a more general meaning.

 At
the LHC energy the diffractive events with the masses as large as
3 TeV could be studied. It would be interesting to check this
prediction at the LHC  where the scaling and simple power-like
behavior of diffraction dissociation differential cross-section
should be observed. Observation of such behavior would confirm
the diffraction mechanism based on excitation of the complex
hadronlike colorless  object - constituent quark. This mechanism
can in principle explain angular structure of diffraction
 in the deep - inelastic
scattering at HERA where smooth angular dependence on the thrust
transverse momentum was observed \cite{will}. If it is the case, then
diffraction in DIS at lower energies should manifest typical soft
diffractive behavior with exponential peak at small $t$ as it does
in hadronic reactions. It would be interesting to check it at
lower energies, e. g. at Jefferson Lab.

 There could be envisaged  various experimental
configurations at the LHC; e.g. soft diffraction  goes well to the interest
of the TOTEM experiment, while hard diffractive final states can
be measured by CMS detector and possible correlations between the
features of the soft and hard diffractive processes can be
obtained using  combined measurements of TOTEM and CMS
\cite{petr}.

 Measurements of the angular dependence of diffraction
dissociation simultaneously  with the measurements of the
differential cross--section of elastic scattering are important
for the determination of  the global geometrical properties of
elastic and inelastic diffraction, their similar and
distinctive features and the origin and nature of the driving
asymptotic mechanism. In general these studies could be an essential
tool to probe non-perturbative QCD.

\section*{Acknowledgements}
We are grateful to V. A. Petrov for the interesting discussions and
comments. This work was supported by RFBR (Grant No. 99-02-17995).

\end{document}